\documentclass[preprint,showpacs,preprintnumbers,natbib]{revtex4}
\usepackage{amssymb}
\usepackage{amsmath}
\usepackage{graphicx}
\usepackage{epsfig}
\usepackage{dcolumn}
\usepackage{bm}

\begin{document}
\title{Relativistic dynamics compels a thermalized Fermi gas
to a unique intrinsic parity eigenstate}
\author{A. E. Bernardini}
\email{alexeb@ufscar.br}
\author{S. S. Mizrahi}
\email{salomon@df.ufscar.br}
\affiliation{Departamento de F\'{\i}sica, Universidade Federal de S\~ao Carlos, PO Box
676, 13565-905, S\~ao Carlos, SP, Brasil}
\date{\today }

\begin{abstract}
Dirac equation describes the dynamics of a relativistic spin-1/2 particle
regarding its spatial motion and intrinsic degrees of freedom. Here we adopt
the point of view that the spinors describe the state of a massive particle 
carrying two qubits of information: helicity and intrinsic
parity. We show that the density matrix for a gas of free fermions, in
thermal equilibrium, correlates helicity and intrinsic parity. Our results
introduce the basic elements for discussing the spin-parity correlation for
a Fermi gas: (1) at the ultra-relativistic domains, when the temperature is
quite high, $T > 10^{10}\ K$, the fermions have no definite intrinsic
parity (50\% : 50\%), which is maximally correlated with the helicity; (2)
at very low temperature, $T \approx 3 \ K$, a unique parity dominates
(conventionally chosen positive), by $10^{20}$ to $1$, while
the helicity goes into a mixed state for spin up and down, and the quantum
correlation decoheres. For the anti-fermions we get the opposite behavior.
In the framework of quantum information, our result could be considered as
a plausible explanation of why we do accept, as a fact (consistent with the
experimental observation), that fermions (and anti-fermions), in our present
epoch of a cool universe, have a unique intrinsic parity. The framework for
constructing spin-parity entangled states is established.
\end{abstract}

\maketitle

\section{Introduction}

Dirac invented his relativistic equation in order to explain the quantum
properties of the electron (spin $1/2$) in the relativistic framework: the equation had
to (\emph{a}) display the formula $E_{p}^{2}=p^{2}+m^{2}$ as the eigenenergy of a particle in
free motion (with $\hbar =c=1$); (\emph{b}) be covariant under a Lorentz
transformation that links the particle dynamical properties between two inertial
frames. Dirac found that the sound equation had to be expressed, necessarily, 
in terms of $4\times 4$ matrices. Its more familiar form is $i\partial \Psi \left( \vec{x}
,t\right) /\partial t=\mathbf{H}_{D}\Psi \left( \vec{x},t\right) $, with the
Hamiltonian being linear in the momentum $\vec{p}$,
\begin{equation}
\mathbf{H}_{D}=\vec{\boldsymbol{\alpha }}\cdot \vec{p}+m\boldsymbol{\beta },
\label{hamdirac}
\end{equation}
and the $4\times 4$ matrices $\vec{\boldsymbol{\alpha }}\equiv (
\boldsymbol{\alpha }_{x},\,\boldsymbol{\alpha }_{y},\,\boldsymbol{\alpha }
_{z})$ and $\boldsymbol{\beta }$, have to satisfy forcefully the relations 
\begin{equation}
\boldsymbol{\alpha }_{k}\boldsymbol{\alpha }_{l}+\boldsymbol{\alpha }_{l}
\boldsymbol{\alpha }_{k}=2I\delta _{kl},\qquad \vec{\boldsymbol{\alpha }}
\boldsymbol{\beta }+\boldsymbol{\beta }\vec{\boldsymbol{\alpha }}=
\boldsymbol{0},\qquad \boldsymbol{\beta }^{2}=\mathbf{I},
\end{equation}
with $\mathbf{I}$ for the unit matrix (in Dirac's book \cite{dirac}, instead
of $\vec{\boldsymbol{\alpha }}$ we find a $4\times 4$ matrix $\boldsymbol{
\rho }_{1}$ multiplying the $4\times 4$ direct product of Pauli matrices $
\vec{\boldsymbol{\sigma }}\equiv (\sigma _{x},\,
\sigma _{y},\,\sigma _{z})$). An usual approach consists in
the introduction of the chiral representation, where the components of the
matrix vector, $\left( \mathbf{I},\vec{\boldsymbol{\sigma }}\right) =
\boldsymbol{\sigma }$ and $\left( \mathbf{I},-\vec{\boldsymbol{\sigma }}
\right) =\boldsymbol{\tilde{\sigma}}$, are respectively in contravariant and
covariant forms, in the same fashion that one has $(x_{\mu })=\left( t,\,-\vec{
x}\right) $ and $(x^{\mu })=\left( t,\,\vec{x}\right) $ \cite{cottingham}. The
state vector solution to the Dirac equation can be written as the sum, 
\begin{equation}
\Psi \left( x\right) =\left( 
\begin{array}{c}
\psi _{L}\left( x\right) \\ 
\mathbf{0}
\end{array}
\right) +\left( 
\begin{array}{c}
\mathbf{0} \\ 
\psi _{R}\left( x\right)
\end{array}
\right) ,  \label{state1}
\end{equation}
of left and right chiral spinors 
\begin{equation*}
\psi _{L}\left( x\right) =\left( 
\begin{array}{c}
\varphi _{L1}\left( x\right) \\ 
\varphi _{L2}\left( x\right)
\end{array}
\right) ,\qquad \psi _{R}\left( t\right) =\left( 
\begin{array}{c}
\chi _{L1}\left( x\right) \\ 
\chi _{L2}\left( x\right)
\end{array}
\right) ,
\end{equation*}
and $x\equiv \left(\vec{x},t\right)$, $\mathbf{0}\equiv \binom{0}{0}$. 
From Dirac equation plus Eq.~(\ref{hamdirac}) and (\ref
{state1}) one constructs two coupled differential equations for the spinors $
\psi _{L}\left( x\right) $ and $\psi _{R}\left( x\right) $, 
\begin{eqnarray*}
i\boldsymbol{\tilde{\sigma}}^{\mu }\partial _{\mu }\psi _{L}\left( x\right)
-m\psi _{R}\left( x\right) &=&0, \\
i\boldsymbol{\sigma }^{\mu }\partial _{\mu }\psi _{R}\left( x\right) -m\psi
_{L}\left( x\right) &=&0,
\end{eqnarray*}
whose Lagrangian is \cite{cottingham} (omitting the explicit dependence on $x$), 
\begin{equation}
\mathcal{L}=i\psi _{L}^{\dagger }\boldsymbol{\tilde{\sigma}}^{\mu }\partial
_{\mu }\psi _{L}+i\psi _{R}^{\dagger }\mathbf{\sigma }^{\mu }\partial _{\mu
}\psi _{R}-m\left( \psi _{L}^{\dagger }\psi _{R}+\psi _{R}^{\dagger }\psi
_{L}\right) .
\end{equation}

Interestingly, the Dirac equation allows a different
insight when written in terms of direct (or Kronecker) products of Pauli
matrices. So, daring to interpret quantum mechanics as a special kind of
information theory for particles and fields \cite{nota2,iwo}, in the
language of quantum information we may
say that the relativistic equation of a spin-1/2 fermion has as solution a
state of two \emph{qubits} (two degrees of freedom)
carried by a massive particle whose dynamical evolution in space is represented by a
continuous variables that may be the position or the linear momentum
\cite{salomon,salomon2}. Hereon we will choose the linear momentum representation (as a c-number)
instead of using the position operator $-i\vec{\nabla}$, since we are not
introducing a position dependent potential in the Hamiltonian. One can
appreciate that fact by writing the matrices $\vec{\boldsymbol{\alpha }}$
and $\boldsymbol{\beta}$ in terms of tensor products of Pauli matrices 
\begin{equation}
\vec{\boldsymbol{\alpha }}=\sigma _{x}^{\left( 1\right)
}\otimes \boldsymbol{\vec{\sigma}}^{\left( 2\right) },\quad \vec{\boldsymbol{
\ \alpha }}\cdot \vec{p}= \sigma _{x}^{\left( 1\right) }\otimes
\left( \vec{p}\cdot \vec{\boldsymbol{\sigma }}^{\left( 2\right) }\right) ,~
\text{and}~~\boldsymbol{\beta }=\sigma _{z}^{\left( 1\right)}\otimes \mathbf{I}_{2},
\end{equation}
where the upperscripts 1 and 2 refer to qubits 1 and 2, respectively. Thus we
write the Dirac Hamiltonian (\ref{hamdirac}) in terms of the direct product of
two-qubit operators, $\mathbf{H}_{D}=\boldsymbol{\sigma }_{x}^{\left( 1\right) }\otimes \left( 
\vec{p}\cdot \vec{\boldsymbol{\sigma }}^{\left( 2\right) }\right) +m \,
\boldsymbol{\sigma }_{z}^{\left( 1\right) }\otimes \mathbf{I}_{2}$,
and the two solutions to Dirac equation are
\begin{eqnarray}
\left\vert \Psi ^{s}(\vec{p},\,t)\right\rangle 
&=&e^{i(-1)^{s}\,E_{p}\,t}\left\vert \psi ^{s}(\vec{p})\right\rangle
= e^{i(-1)^{s}\,E_{p}\,t}N_{s}\left( p\right)  \notag
\\
&\times& \left[ \left\vert
+\right\rangle _{1}\otimes \left\vert u(\vec{p})\right\rangle _{2}+\left( 
\frac{p}{E_{p}+(-1)^{s+1}m}\right) |-\rangle _{1}\,\otimes \left( \hat{p}
\cdot \vec{\boldsymbol{\sigma }}^{\left( 2\right) }\left\vert u(\vec{p}
)\right\rangle _{2}\right) \right] , \label{sol1}
\end{eqnarray}
where $s=0$ and $1$ stand respectively for negative and positive energy solutions, 
$\vec{p}=p\,\hat{p}$, with $\left\vert \hat{p}\right\vert = 1$. 
The state $\left\vert u(\vec{p})\right\rangle _{2}$ is a spinor
representing the spatial motion of the free fermion ($u(\vec{p})$ in the
momentum representation) coupled to its spin, which describes a
structureless magnetic dipole moment. For qubit 1 the kets, $\left\vert +\right\rangle _{1}$ 
and $ |-\rangle _{1}$, are identified as the intrinsic parity eigenstates of the
fermion. The states are orthogonal, $\left\langle \pm |\pm (\mp )\right\rangle _{1}=1(0)$.
For the inner product we get $\left\langle \Psi ^{s}(\vec{p},\,t)|\Psi ^{s}(\vec{
p},\,t)\right\rangle =\left\langle u(\vec{p})|u(\vec{p})\right\rangle _{2}$,
with the normalization factor 
\begin{equation}
N_{s}(p)=\frac{1}{\sqrt{2}}\left( 1+(-1)^{s+1}\frac{m}{E_{p}}\right) ^{1/2},
\label{norm}
\end{equation}
and we also assume that the local probability distribution for
the momenta is normalized, $\int {d^{3}p\, \left\langle 
u(\vec{p})|u(\vec{p})\right\rangle}_{2}=1$.
Thus the spinors and $4\times 4$ matrices stand for the the direct product of 
the intrinsic degrees of freedom of a massive spin-1/2 fermion, 
parametrized by the linear momentum 
$\vec{p}$, on free motion in space. Since $\mathbf{H}_{D}\left\vert \psi
^{s}\left( \vec{p}\right) \right\rangle =(-1)^{s+1}E_{p}\left\vert \psi
^{s}\left( \vec{p}\right) \right\rangle $, one has $\left( \mathbf{H}
_{D}\right) ^{2}\left\vert \psi ^{s}\left( \vec{p}\right) \right\rangle
=E_{p}^{2}\left\vert \psi ^{s}\left( \vec{p}\right) \right\rangle $ that
leads to Einstein's dispersion relation $p^{2}+m^{2}=E_{p}^{2}$. As so, the
state (\ref{sol1}) has no definite intrinsic parity, qubit 1
is in a superposition of both eigenstates. 

The total parity operator $\hat{P}$ acts on the Kronecker product $
\left\vert \pm \right\rangle _{1}\otimes \left\vert u(\vec{p})\right\rangle
_{2}$ as $\hat{P}\left( \left\vert \pm \right\rangle _{1}\otimes \left\vert
u(\vec{p})\right\rangle _{2}\right) =\pm \left( \left\vert \pm \right\rangle
_{1}\otimes \left\vert u(-\vec{p})\right\rangle _{2}\right) $; indeed it is
the product of two operators, the intrinsic parity $\hat{P}^{int}$ (having
two eigenvalues, $\hat{P}^{int}\left\vert \pm \right\rangle =\pm \left\vert
\pm \right\rangle $) and the spatial parity $\hat{P}^{sp}$ ($\hat{P}
^{sp}\varphi \left( \vec{p}\right) =\varphi \left( -\vec{p}\right) $). Thus,
$\hat{P}^{int}=\beta =\sigma _{z}^{\left( 1\right)}\otimes I^{\left( 2\right) }$
applies on $\left\vert \Psi ^{s}(\vec{p},\,t)\right\rangle $, Eq. (\ref{sol1}
), and  it follows that $ \hat{P} ^{-1}=\hat{P}$. Regarding the spatial
parity operator 
\begin{equation*}
\hat{P}^{sp}\left\{ 
\begin{array}{c}
\vec{r} \\ 
\vec{p}
\end{array}
\right\} \hat{P}^{sp}=-\left\{ 
\begin{array}{c}
\vec{r} \\ 
\vec{p}
\end{array}
\right\} \text{,\quad }\hat{P}^{sp}\left\{ 
\begin{array}{c}
\vec{l} \\ 
\vec{\sigma}
\end{array}
\right\} \hat{P}^{sp}= + \left\{ 
\begin{array}{c}
\vec{l} \\ 
\vec{\sigma}
\end{array}
\right\} ,
\end{equation*}
the $+$ ($-$) sign stands for axial (polar) vectors. Complementarily, the $
\gamma$-matrices are $\gamma ^{0}=\beta =\sigma _{z}^{\left( 1\right)
}\otimes I^{\left( 2\right) }$, $\gamma ^{i}=i\sigma _{y}\otimes \sigma _{i}$
, $i=1,2,3$, and $\gamma ^{5}=i\gamma ^{0}\gamma ^{1}\gamma ^{2}\gamma
^{3}=\sigma _{x}^{\left( 1\right) }\otimes I^{\left( 2\right) }$. 

\section{Correlation between the intrinsic degrees of freedom}

There is an asymmetry between the two terms within the brackets
in the solution (\ref{sol1}): the first one represents the limit for the
non-relativistic state of a spin $1/2$ free fermion, namely, the solution to
the Schr\"{o}dinger equation, while the second term is responsible for the
relativistic effect (containing the helicity operator $\hat{p}\cdot \vec{ 
\boldsymbol{\sigma }}_{2}$). Due to the correlation between the
parity and helicity qubits, a hypothetical measurement that results in qubit 
$\left\vert +\right\rangle _{1}$ would reduces
the Dirac solution to the non-relativistic Schr\"{o}dinger equation while if
the result of the measurement is qubit $\left\vert -\right\rangle _{1}$
the solution is reduced to the purely relativistic term. However,  
there is no hint that the nature selects one of the two 
components under any kind of measurement. Nevertheless, as we are going to see 
below, for an ensemble of fermions in contact with a thermal reservoir,
one intrinsic parity eigenstate will be selected naturally as an effect 
of cooling. 

The helicity eigenvalue equation $\hat{p}\cdot 
\vec{\boldsymbol{\sigma }}\,\left\vert \Omega _{\pm }\right\rangle =\pm
\left\vert \Omega _{\pm }\right\rangle $ has orthogonal eigenstates 
\begin{eqnarray*}
\left\vert \Omega _{+}\right\rangle &=&\cos \left( \theta /2\right)
\left\vert \uparrow \right\rangle +e^{i\phi }\sin \left( \theta /2\right)
\left\vert \downarrow \right\rangle , \\
\left\vert \Omega _{-}\right\rangle &=&\sin \left( \theta /2\right)
\left\vert \uparrow \right\rangle -e^{i\phi }\cos \left( \theta /2\right)
\left\vert \downarrow \right\rangle ,
\end{eqnarray*}
($\left\langle \Omega _{+}|\Omega _{+}\right\rangle =\left\langle \Omega
_{-}|\Omega _{-}\right\rangle =1$, $\left\langle \Omega _{+}|\Omega
_{-}\right\rangle =0$) where the angles $\theta $ and $\phi $ determine
the direction of $\hat{p}$ (on a spherical surface of radius 1 the tips of the
versors $\hat{p}$ and $-\hat{p}$ are localized by the angles $\Omega
_{+}\equiv \left( \theta ,\phi \right) $ and $\Omega _{-}\equiv \left( \pi
-\theta ,\phi +\pi \right) $) and the kets $\left\vert \uparrow
\right\rangle $, $\left\vert \downarrow \right\rangle $ stand for $\binom{1}{
0 }$ and $\binom{0}{1}$. So, the spinor $\left\vert u(\vec{p})\right\rangle
_{2}$ can be written as the superposition 
\begin{equation}
\left\vert u(\vec{p})\right\rangle =A(\vec{p})\left\vert \Omega
_{+}\right\rangle +B(\vec{p})\left\vert \Omega _{-}\right\rangle ,  \label{h}
\end{equation}
(we omit the subscript $2$) where $|A(\vec{p})|^{2}+|B(\vec{p})|^{2}$ is the
density distribution of the linear momentum. The spinor (\ref{h}) correlates
the linear momentum (a c-number) to the helicity eigenstates, however,
for simplicity, we are going to assume that the linear momentum is not correlated to
the helicity, therefore 
\begin{equation}
A(\vec{p})=\varphi (\vec{p})\cos {(\chi )}\,,\quad B(\vec{p})=\varphi (\vec{
p })e^{i\mu }\sin {(\chi )},
\end{equation}
where a mixing angle, ${\chi \in \left[0,\pi\right] }$, and a relative phase,
$\mu \in \left[0,2\pi \right) $, have been introduced. The helicity sector 
of the Dirac equation solution will make use of the spinors 
\begin{equation}
\left\vert u_{\pm }(\vec{p})\right\rangle =\varphi (\vec{p})\left\vert
h_{\pm }\right\rangle ,
\end{equation}
with 
\begin{eqnarray}
\left\vert h_{+}\right\rangle &\equiv &\left\vert h\right\rangle =\cos {\
(\chi )}\,\left\vert \Omega _{+}\right\rangle +e^{i\mu }\,\sin {(\chi )}
\left\vert \Omega _{-}\right\rangle ,  \notag \\
\left\vert h_{-}\right\rangle &\equiv &\hat{p}\cdot \vec{\boldsymbol{\sigma }
}\,\left\vert h\right\rangle =\cos {(\chi )}\,\left\vert \Omega
_{+}\right\rangle -e^{i\mu }\,\sin {(\chi )}\left\vert \Omega
_{-}\right\rangle ,  \label{polar}
\end{eqnarray}
that are normalized $\left\langle h_{\pm }|h_{\pm }\right\rangle =1$,
however they are orthogonal only for $\chi =\pi /4$, because $\left\langle
h_{+}|h_{-}\right\rangle =\cos {\left( 2\chi \right) }$. It is worth noting 
that doing the changes $\chi\rightarrow \pi - \chi$ and $\phi \rightarrow \phi + \pi$ 
we get $\left\vert h_{+}\right\rangle \rightarrow -\left\vert h_{-}\right\rangle$.
For a normalized linear momentum distribution, $\int {d^{3}p\,\left\vert \varphi (\vec{p}
)\right\vert ^{2}}=1$, one has 
\begin{equation}
\int {d^{3}p\,\left\langle u_{\pm }(\vec{p})|u_{\pm }(\vec{p})\right\rangle }
=1,
\end{equation}
and 
\begin{equation}
\int d^{3}p\ \left\langle u_{+}(\vec{p})|u_{-}(\vec{p})\right\rangle =\cos {
\ (2\chi )}.
\end{equation}

The simplified form of the time-independent component of Eq.~(\ref{sol1}) becomes 
\begin{equation}
\left\vert \psi ^{\left( s\right) }(\vec{p})\right\rangle \equiv \varphi ( 
\vec{p})\left\vert \eta _{s}(p)\right\rangle ,
\end{equation}
where 
\begin{equation}
\left\vert \eta _{s}(p)\right\rangle =N_{s}(p)\left( \left\vert
+\right\rangle _{1}\otimes \left\vert h_{+}\right\rangle _{2}+\frac{p}{
E_{p}+(-1)^{s+1}m}\left\vert -\right\rangle _{1}\otimes \left\vert
h_{-}\right\rangle _{2}\right) ,
\end{equation}

with $\left\langle \eta _{s}(p)|\eta _{s}(p)\right\rangle =1$, and the pure
state density matrix is 
\begin{equation}
\rho _{12}^{\left( s\right) }(\vec{p})=\left\vert \psi ^{\left( s\right) }( 
\vec{p})\right\rangle \left\langle \psi ^{\left( s\right) }(\vec{p}
)\right\vert =\left\vert \varphi (\vec{p})\right\vert ^{2}\left\vert \eta
_{s}(p)\right\rangle \,\left\langle \eta _{s}(p)\right\vert .  \label{pure}
\end{equation}

Calculating the trace over the qubits the result is $ \mathrm{Tr}_{12}\left[
\rho _{12}^{\left( s\right) }\left(\vec{p}\right)\right] = \left\vert \varphi (\vec{p}
)\right\vert ^{2}$, thus $\int {d^{3}p \,\mathrm{Tr}_{12}\left[
\rho_{12}^{\left( s\right) }(\vec{p})\right] }=1 $. 

For an ensemble of free fermions interacting with a thermal environment at 
temperature $T$, we identify the probability density $\left| \varphi (\vec{p}
)\right|^{2}$ with a normalized distribution function isotropic in 
the linear momentum, $\int {d^{3}p\,f(p,\,T)}=1$. Integrating Eq.~(\ref{pure}) 
over the linear momentum, the reduced density matrix becomes
\begin{equation}
\rho _{12}^{\left( s\right) }=\int_{0}^{\infty }{dp\,p^{2}\,f(p,T)\,\left(
\int {d}\Omega {\,\left\vert \eta _{s}(p)\right\rangle \left\langle \eta
_{s}(p)\right\vert }\right) },  \label{rho3}
\end{equation}
where we omit the subscripts $1$ and $2$ in the right-side. 
As the dependence on the solid angle $\Omega$ is exclusively
relegated to the helicity states $\left\vert \Omega _{\pm}\right\rangle$, we
get 
\begin{eqnarray}
\int {d}\Omega \left\vert \Omega _{\pm }\right\rangle \left\langle \Omega
_{\pm }\right\vert {\,} &=&\frac{1}{2}\mathbf{I}\left( \int {d}\Omega
\right) ,  \notag \\
\int {d}\Omega {\,\left\vert \Omega _{\pm }\right\rangle \left\langle \Omega
_{\mp }\right\vert } &=&\frac{\pi }{8}\mathbf{\sigma }_{z}\left( \int {d}
\Omega \right) ,
\end{eqnarray}
where $\mathbf{I}= \left\vert \uparrow \rangle \langle
\uparrow \right\vert ~+~ \left\vert \downarrow \rangle \langle \downarrow \right\vert $ 
and $\mathbf{\sigma}_{z} = \left\vert \uparrow \rangle \langle \uparrow \right\vert
~-~\left\vert \downarrow \rangle \langle\downarrow \right\vert$ 
from which, by Eq.~(\ref{polar}), one obtains 
\begin{equation*}
\left( \int {d}\Omega \right) ^{-1}\int {d}\Omega {\,|h_{\pm }\rangle
\langle h_{\pm }|} = n_{\pm }\left\vert \uparrow
\right\rangle \left\langle \uparrow \right\vert +n_{\mp }\left\vert
\downarrow \right\rangle \left\langle \downarrow \right\vert ,
\end{equation*}
with real coefficients 
\begin{equation*}
n_{\pm }=\frac{1}{2}\pm \frac{\pi }{8}\sin {\left( 2\chi \right) }\,
\cos {\ (\mu )},
\end{equation*}
and 
\begin{equation*}
\left( \int {d}\Omega \right) ^{-1}\int {d}\Omega {\,|h_{\pm }\rangle
\langle h_{\mp }|}= \tilde{n}
_{\mp }\left\vert \uparrow \right\rangle \left\langle \uparrow \right\vert + 
\tilde{n}_{\pm }\left\vert \downarrow \right\rangle \left\langle \downarrow
\right\vert ,
\end{equation*}

with complex coefficients 

\begin{equation*}
\tilde{n}_{\pm }=\frac{1}{2}\cos {(2\chi )}\pm \frac{i\pi }{8}\sin
\left(2\chi \right) \sin {(\mu )},
\end{equation*}
noting that $n_{+}+n_{-}=1$, $\tilde{n}_{+}+\tilde{n}_{-}=\cos {(2\chi )}$
and $\left( \tilde{n}_{-}\right) ^{\ast }=\tilde{n}_{+}$. 

The reduced density operator (\ref{rho3}) becomes 
\begin{eqnarray}
\hat{\rho}_{12}^{\left( s\right) } &=&{M}_{++}^{s}\left( T\right) \left[
\left\vert +\right\rangle \left\langle +\right\vert \otimes \left(
n_{+}\left\vert \uparrow \right\rangle \left\langle \uparrow \right\vert
+n_{-}\left\vert \downarrow \right\rangle \left\langle \downarrow
\right\vert \right) \right]  \notag \\
&&+{M}_{--}^{s}\left( T\right) \left[ \left\vert -\right\rangle \left\langle
-\right\vert \otimes \left( n_{-}\left\vert \uparrow \right\rangle
\left\langle \uparrow \right\vert +n_{+}\left\vert \downarrow \right\rangle
\left\langle \downarrow \right\vert \right) \right]  \notag \\
&&+{M}_{+-}^{s}\left( T\right) \left[ \left\vert +\right\rangle \left\langle
-\right\vert \otimes \left( \tilde{n}_{-}\left\vert \uparrow \right\rangle
\left\langle \uparrow \right\vert +\tilde{n}_{+}\left\vert \downarrow
\right\rangle \left\langle \downarrow \right\vert \right) \right. ,  \notag
\\
&&+\left. \left\vert -\right\rangle \left\langle +\right\vert \otimes \left( 
\tilde{n}_{+}\left\vert \uparrow \right\rangle \left\langle \uparrow
\right\vert +\tilde{n}_{-}\left\vert \downarrow \right\rangle \left\langle
\downarrow \right\vert \right) \right],  \label{dens4}
\end{eqnarray}
where we defined the coefficients
\begin{eqnarray}
M_{++}^{s}(T) &=& \int{d^{3}p\,f(p,T)\,\,N_{s}^{2}(p)},  \notag \\
M_{--}^{s}(T) &=&\int{d^{3}p\,f(p,T)\,\,N_{s}^{2}(p)\,g^2_{s}\left(
p,m\right)} ,  \notag \\
M_{+-}^{s}(T) &=& \int {d^{3}p\,f(p,T)\,\,N_{s}^{2}(p)\,g_{s}\left(
p,m\right)} ,  \label{M+-}
\end{eqnarray}
with ${M}_{++}^{s}\left( T \right) + {M}_{--}^{s}\left( T \right) =1$, 
${N_{s}^{2}(p)}\left( 1+ g_{s}^{2}\left( p,m \right) \right) =1$,
and $g_{s}\left( p,m \right) = p/ \left( E_{p}+(-1)^{s+1}m \right) $. 

As we admitted that qubit 1 stands for the intrinsic parity of the fermion,
we get the correlation density operator for helicity and
intrinsic parity as suggested many years ago by T. D. Lee and C. N. Yang
\cite{leeyang}, in an \emph{ad-hoc} procedure.
In that case (c. f. Eq.~(23) from Ref.~\cite{leeyang}) the density matrix describes a coherent collection of {\em spinorial} particles that exhibit spin and parity as correlated quantum features, through which a suitable interference phenomena between parity doublets is identified.
Furthermore, setting $q=p/T$
the Fermi-Dirac distribution can be written as 
\cite{Dodelson,Ma94,Bernardini2011,Bernardini2011B} 
\begin{equation}
f(p,T)=\frac{1}{6\pi \zeta (3)}\frac{1}{T^{3}}\left( e^{q}+1 \right) ^{-1},
\end{equation}
where we have set the Boltzmann constant $k = 1$ and $\zeta(3)\approx 1.202$ 
is a Riemann zeta function. 
The coefficients (\ref{M+-})) can be calculated numerically,
\begin{eqnarray}
M_{++}^{s}(T_{m}) &=&\frac{1}{3\zeta (3)}
\int_{0}^{\infty}{\ dq\,q^{2}\,\left( 1+(-1)^{s+1}\frac{1}{\sqrt{
1+T_{m}^{2}\,q^{2}}}\right) \, \frac{1}{e^{q}+1}},  \notag \\
M_{--}^{s}(T_{m}) &=&\frac{1}{3\zeta (3)}
\int_{0}^{\infty}{\ dq\,q^{2}\,\left( 1+(-1)^{s}\frac{1}{\sqrt{
1+T_{m}^{2}\,q^{2}}}\right) \, \frac{1}{e^{q}+1}},  \label{matrix} \\
M_{+-}^{s}(T_{m}) &=&\frac{1}{3\zeta (3)}
\int_{0}^{\infty}{dq\, \frac{T_{m}\,q^{3}}{\sqrt{1+T_{m}^{2}\,q^{2}}}\ \frac{1
}{e^{q}+1}},  \notag
\end{eqnarray}
with $kT/mc^{2}\longrightarrow T_{m}=T/m$ (temperature per unit mass with $k=c=1$). 
In Table I we present the values of the $M_{ij}^{s} $ for different 
temperatures. 
\begin{table}
\begin{tabular}{c|c|c|c}
\hline\hline
$T_{m}$ & $ M_{++}^{1} (T_m )= M_{--}^{0} 
(T_{m} )$ & $ M_{--}^{1} (T_{m} )= M_{++}^{0}(T_{m}) $ & $ M_{+-}^{1} =M
_{+-}^{0}$ \\ \hline\hline
\multicolumn{1}{l|}{${ 10}^{5}$} & \multicolumn{1}{|l|}{${ 
0.500\,00}$} & \multicolumn{1}{|l|}{${ 0.500\,00}$} & 
\multicolumn{1}{|l}{${ 0.500\,00}$} \\ 
\multicolumn{1}{l|}{${ 10}^{2}$} & \multicolumn{1}{|l|}{${ 
0.502\,28}$} & \multicolumn{1}{|l|}{${ 0.497\,72}$} & 
\multicolumn{1}{|l}{${ 0.499\,99}$} \\ 
\multicolumn{1}{l|}{${ 10}$} & \multicolumn{1}{|l|}{${ 
0.522\,64}$} & \multicolumn{1}{|l|}{${ 0.477\,36}$} & 
\multicolumn{1}{|l}{${ 0.499\,12}$} \\ 
\multicolumn{1}{l|}{${ 1}$} & \multicolumn{1}{|l|}{${ 
0.685\,87}$} & \multicolumn{1}{|l|}{${ 0.314\,13}$} & 
\multicolumn{1}{|l}{${ 0.452\,46}$} \\ 
\multicolumn{1}{l|}{${ 10}^{-1}$} & \multicolumn{1}{|l|}{${ 
0.972\,98}$} & \multicolumn{1}{|l|}{${ 2.\,\allowbreak 702\,1\times
10}^{-2}$} & \multicolumn{1}{|l}{${ 0.144\,65}$} \\ 
\multicolumn{1}{l|}{${ 10}^{-2}$} & \multicolumn{1}{|l|}{${ 
0.999\,68}$} & \multicolumn{1}{|l|}{${ 3.\,\allowbreak 227\,5\times
10}^{-4}$} & \multicolumn{1}{|l}{${ 1.\,\allowbreak 574\,1\times 10}
^{-2}$} \\ 
\multicolumn{1}{l|}{${ 10}^{-5}$} & \multicolumn{1}{|l|}{${ 
1.\,\allowbreak 000\,0}$} & \multicolumn{1}{|l|}{${ 3.\,\allowbreak
234\,9\times 10}^{-10}$} & \multicolumn{1}{|l}{${ 1.\,\allowbreak
575\,7\times 10}^{-5}$} \\ 
\multicolumn{1}{l|}{${ 10}^{-10}$} & \multicolumn{1}{|l|}{${ 
1.0000}$} & \multicolumn{1}{|l|}{${ 3.\,\allowbreak 234\,9\times 10}
^{-20}$} & \multicolumn{1}{|l}{${ 1.\,\allowbreak 575\,7\times 10}
^{-10}$} \\ 
\multicolumn{1}{l|}{${ 10}^{-12}$} & \multicolumn{1}{|l|}{${ 
1.0000}$} & \multicolumn{1}{|l|}{${ 3.\,\allowbreak 234\,9\times 10}
^{-24}$} & \multicolumn{1}{|l}{${ 1.\,\allowbreak 575\,7\times 10}
^{-12}$} \\ \hline\hline
\end{tabular}
\caption{\small{Temperature per unit mass and the values of the
coefficients in Eqs. (\ref{matrix}).}}
\end{table}

While at very high temperature $T_{m}> { 10}^{2}$ (for electrons 
\footnote{$mc^{2}=5.11\times 10^{5} \, eV$ and $k=8.617 \,\times 10^{-5}\, eV/K $},
it corresponds to $T > 10^{12} \, K$) we have $ M_{++}^{s}
(T_{m}) \approx M_{--}^{s}(T_{m})\approx M_{+-}^{s} \left( T
_{m}\right) \approx  0.5 $, the coefficients take, nearly, the same
values independently of $s$; the gas has an equilibrated distribution of
intrinsic parity for fermions ($s=1$) and for antifermions ($s=0$). As the
temperature reduces, the distributions change. For instance, at $T_{m}= 
{ 10}^{-2}$ ($T=10^{8}\,K$) and $s=1$, the gas is constituted,
overwhelmingly, by positive parity fermions, while for $s=0$ it is,
mostly, made of negative parity antifermions. The transition
probabilities $ M_{+-}^{1}$ and $ M_{+-}^{0}$ are the same
at any temperature, and they vanish as the gas cools down. In Table II we
give the differences between the $M_{ij}^{s}$ 
 
\begin{table}
\begin{tabular}{c|c|c}
\hline\hline
$ T_{m}$ & $ M_{++}^{1} - M_{--}^{1}$ & $ M
_{++}^{0} - M_{--}^{0}$ \\ \hline\hline
\multicolumn{1}{l|}{${ 10}^{5}$} & \multicolumn{1}{|l|}{${ 0}$
} & \multicolumn{1}{|l}{${ 0}$} \\ 
\multicolumn{1}{l|}{${ 10}^{2}$} & \multicolumn{1}{|l|}{${ 
0.004\,56}$} & \multicolumn{1}{|l}{${ -0.004\,56}$} \\ 
\multicolumn{1}{l|}{${ 10}$} & \multicolumn{1}{|l|}{${ 
0.045\,28}$} & \multicolumn{1}{|l}{${ -0.045\,28}$} \\ 
\multicolumn{1}{l|}{${ 1}$} & \multicolumn{1}{|l|}{${ 
0.371\,74}$} & \multicolumn{1}{|l}{${ -0.371\,74}$} \\ 
\multicolumn{1}{l|}{${ 10}^{-1}$} & \multicolumn{1}{|l|}{${ 
0.945\,96}$} & \multicolumn{1}{|l}{${ -0.945\,96}$} \\ 
\multicolumn{1}{l|}{${ 10}^{-2}$} & \multicolumn{1}{|l|}{${ 
0.999\,36}$} & \multicolumn{1}{|l}{${ -0.999\,36}$} \\ 
\multicolumn{1}{l|}{$ \leq { 10}^{-5}$} & \multicolumn{1}{|l|}{${ 1}
$} & \multicolumn{1}{|l}{${ -1}$} \\ 
\hline\hline
\end{tabular}
\caption{\small{Temperature per unit mass and the difference between 
coefficients in Eqs. (\ref{matrix}).}}
\end{table}

\begin{figure}[tbp]
{\normalsize \includegraphics{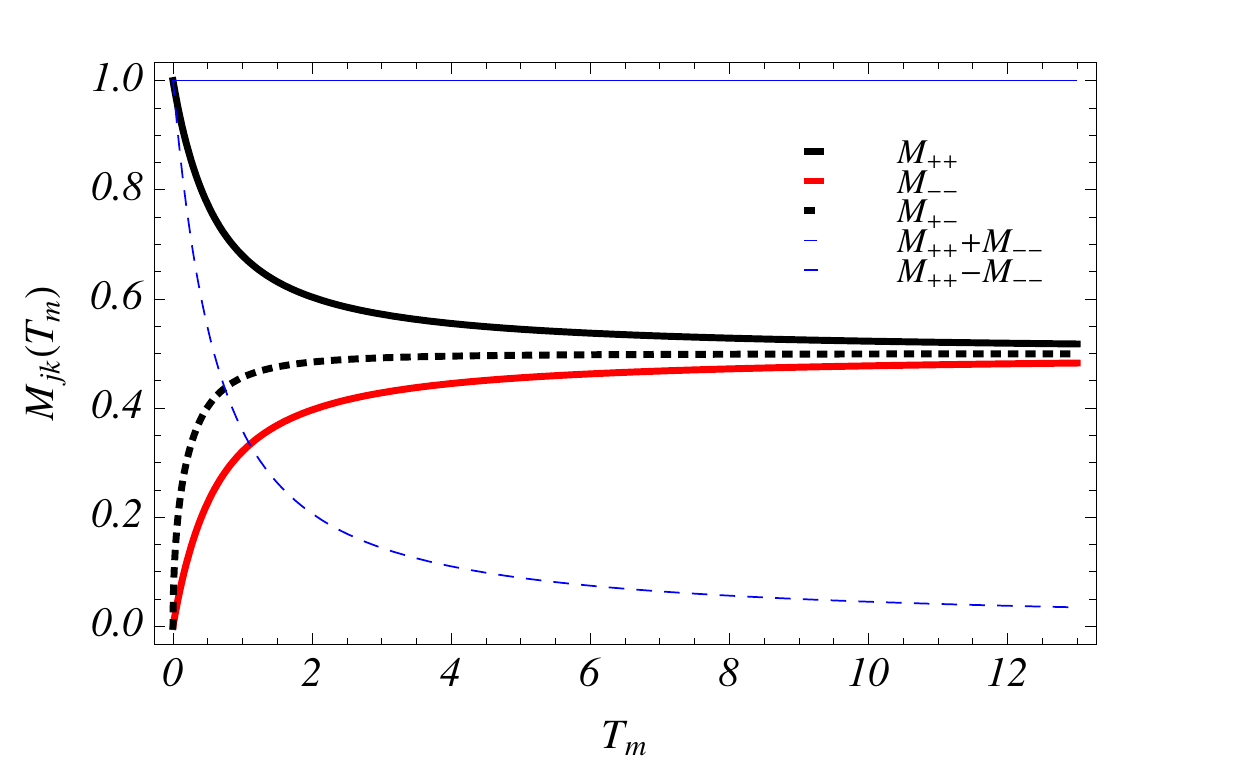}}
\caption{Density matrix coefficients $M_{ij}$ as a function of the
temperature parameter $T_{m}$.}
\label{Fig01}
\end{figure}

In Fig.~\ref{Fig01} we have drawn the coefficients $M_{ij}^{1}$ (for fermions)
as function of $T_{m}$, that contain the amount of correlation between
spin and intrinsic parity 
of a Fermi gas embedded in a thermalized environment. 
Now we speculate about the meaning of the results. Focusing our attention
on the cosmological scenario, the temperature of the universe is a parameter
that, roughly, parallels the evolution in time: after the initial surge of a very hot
and compact seed of energy, the radiation dominated universe expands and cools down,
still keeping the thermalized blackbody frequency distribution. The
higher (lower) the temperature the earlier (later) is its age and size, 
$T\propto 1/a(t)$ where $a(t)$ is the expansion parameter. At quite high
temperatures, $T_{m}\gg 1$, or $T\gg mc^{2}/k \,$ (for electrons $
T\gg6\times 10^{9} \ K$), the coefficients $M_{++}^{1}$, $M_{--}^{1}$ and $
M_{+-}^{1}$ are close to $0.5$, while as $T_{m}\longrightarrow 0$, $
M_{++}^{1}\lesssim 1$ and $M_{--}^{1}$, and $M_{+-}^{1}$ go to zero. Thus, at
early times, when the universe was quite hot, the fermions (and also antifermions) 
existed with positive
and negative intrinsic parity ($M_{++}^{1}\approx $ $M_{--}^{1}\approx 0.5$),
whereas the transition amplitudes, from positive to negative parity, and
vice versa, were almost the same. As the universe gone expanding
the temperature was reducing, then one parity (positive) began to dominate over the other,
$M_{++}^{1}\longrightarrow 1$, $
M_{--}^{1}\longrightarrow 0$), and the transition amplitude $
M_{+-}^{1}\longrightarrow 0$ was reducing too, so the negative parity fermions
became scarcer. As the universe cooled further, say at $T_{m}=3\times { 10}
^{-10}$ ($3 \, K$) for $s=1$, the fraction of negative to positive parity
fermions became $10^{-20}$, while the inverse comes out for the $s=0$
antifermions. At the present epoch the positive parity fermions dominate,
constituting the building blocks of the observed universe, while those having
negative parity are reduced to almost beyond observation. 

By its turns if we consider the negative energy solution ($s=0$
), for the antifermions, the inverse occurs, negative parity prevails at low
temperature while the positive parity fermions become quite scarce. 
So after our calculations the separation between positive parity fermions 
and negative parity antifermions in the present cold universe finds a plausible explanation.

\section{Charge conjugation}

The charge conjugation operation changes matter into antimatter
and it is represented by the operator is $\hat{C}=-i\gamma ^{2}K\equiv
\left( \sigma_{y}^{\left( 1\right) }\otimes \sigma _{y}^{\left( 2\right)
}\right) \hat{K}$, where $\hat{K}$ stands for the complex conjugation
operator and $\hat{C}^{-1}=-\hat{C}$. It transforms a Dirac spinor as
\begin{equation*}
\left( 
\begin{array}{c}
\left\vert \Psi _{L}^{C}\left( \vec{p},t\right) \right\rangle \\ 
\left\vert \Psi _{R}^{C}\left( \vec{p},t\right) \right\rangle
\end{array}
\right) \equiv \hat{C}\left( 
\begin{array}{c}
\left\vert \Psi _{L}\left( \vec{p}.t\right) \right\rangle \\ 
\left\vert \Psi _{R}\left( \vec{p},t\right) \right\rangle
\end{array}
\right) =\left( 
\begin{array}{c}
-i\sigma _{y}^{\left( 2\right) }\left\vert \Psi _{R}^{\ast }\left( \vec{p}
,t\right) \right\rangle \\ 
i\sigma _{y}^{\left( 2\right) }\left\vert \Psi _{L}^{\ast }\left( \vec{p}
,t\right) \right\rangle
\end{array}
\right).
\end{equation*}

Thus a state is invariant under $\hat{C}$ operation whenever $\left\vert
\Psi _{L}^{C}\left( \vec{p},t\right) \right\rangle =-i\sigma _{y}^{\left(
2\right) }\left\vert \Psi _{R}^{\ast }\left( \vec{p},t\right) \right\rangle $
and $\left\vert \Psi _{R}^{C}\left( \vec{p},t\right) \right\rangle =i\sigma
_{y}^{\left( 2\right) }\left\vert \Psi _{L}^{\ast }\left( \vec{p},t\right)
\right\rangle $. Applying $\hat{C}$ on (\ref{sol1}) one obtains 
\begin{eqnarray*}
\left( 
\begin{array}{c}
\left\vert \Psi _{L}^{C}\left( \vec{p}\right) \right\rangle \\ 
\left\vert \Psi _{R}^{C}\left( \vec{p}\right) \right\rangle
\end{array}
\right) &=&\hat{C}\left( 
\begin{array}{c}
\left\vert u(\vec{p})\right\rangle _{2} \\ 
\left( \frac{p}{E_{p}+(-1)^{s}m}\right) \left( \hat{p}\cdot \vec{\boldsymbol{
\ \sigma }}_{2}\left\vert u(\vec{p})\right\rangle _{2}\right)
\end{array}
\right) \\
&=&\left( 
\begin{array}{c}
-\left( \frac{p}{E_{p}+(-1)^{s+1}m}\right) i\sigma _{y}^{\left( 2\right)
}\left( \hat{p}\cdot \vec{\boldsymbol{\sigma }_{2}^{\ast }}\left\vert
u^{\ast }(\vec{p})\right\rangle _{2}\right) \\ 
i\sigma _{y}^{\left( 2\right) }\left\vert u^{\ast }(\vec{p})\right\rangle
_{2}
\end{array}
\right) ,
\end{eqnarray*}
implying also the change $s\longrightarrow s+1$. Now, applying $\hat{C}
^{-1 }$ on the right and $\hat{C}$ on the left of state (\ref{dens4}),
the coefficients do not change, while $\left\vert \pm
\right\rangle\longrightarrow \left\vert \mp \right\rangle $ and $\left\vert
\uparrow \downarrow \right\rangle \longrightarrow \left\vert \downarrow
\uparrow \right\rangle $, thus $\hat{C}\hat{\rho}_{12}^{\left( s\right) }
\hat{C}^{-1 }$ differs from $\hat{\rho}_{12}^{\left( s\right) }$ by the
following interchanges ${M}_{++}^{\left( s+1\right) }\left(
T\right)\rightleftarrows {M}_{--}^{\left( s+1\right) }\left( T\right) $ and $
\tilde{n}_{+}\rightleftarrows \tilde{n}_{-}$, or, fermion ($\hat{\rho}
_{12}^{\left(1\right) }$) and antifermion ($\hat{C}\hat{\rho}_{12}^{\left(
0\right) }\hat{C}^{-1 }$) interchange their role. 

\section{The density matrices}

The parity-helicity density matrix is 
\begin{equation}
\hat{\rho}_{12}^{\left( s\right) }=\left( 
\begin{array}{cccc}
n_{+}{M}_{++}^{s}\left( T_{m}\right) & 0 & \tilde{n}_{-}{M}_{+-}^{s}\left(
T_{m}\right) & 0 \\ 
0 & n_{-}{M}_{++}^{s}\left( T_{m}\right) & 0 & \tilde{n}_{+}{M}
_{+-}^{s}\left( T_{m}\right) \\ 
\tilde{n}_{+}{M}_{+-}^{s}\left( T_{m}\right) & 0 & n_{-}{M}_{--}^{s}\left(
T\right) & 0 \\ 
0 & \tilde{n}_{-}{M}_{+-}^{s}\left( T_{m}\right) & 0 & n_{+}{M}
_{--}^{s}\left( T_{m}\right)
\end{array}
\right)  \label{dens5}
\end{equation}
from which we verify that, under the Peres-Horodecki criterion
\cite{peres,horo}, there is no entanglement between intrinsic parity and
helicity since the partially transposed matrix, $\left( \hat{1} \times \hat{T
}\right) \hat{\rho}_{12}^{\left( s\right)}$, coincides with $\hat{\rho}
_{12}^{\left( s\right) }$. The eigenvalues of $\hat{\rho}_{12}^{\left(
s\right) }$ are 
\begin{eqnarray*}
\lambda _{1}^{\left( s\right) } &=&\frac{1}{2}\left( n_{-}{M}_{++}^{s} 
+n_{+}{M}_{--}^{s} \right)+\frac{1}{2}\sqrt{\left( n_{-}{M}_{++}^{s} -n_{+}{M}_{--}^{s} \right)
^{2}+4\tilde{n}_{+}\tilde{n}_{-}\left( {M}_{+-}^{s} \right)
^{2}}, \\
\lambda _{2}^{\left( s\right) } &=&\frac{1}{2}\left( n_{-}{M}_{++}^{s} 
+n_{+}{M}_{--}^{s} \right)-\frac{1}{2}\sqrt{\left( n_{-}{M}_{++}^{s} -n_{+}{M}_{--}^{s} \right)
^{2}+4\tilde{n}_{+}\tilde{n}_{-}\left( {M}_{+-}^{s} \right)
^{2}}, \\
\lambda _{3}^{\left( s\right) } &=&\frac{1}{2}\left( n_{+}{M}_{++}^{s}
+n_{-}{M}_{--}^{s} \right)+\frac{1}{2}\sqrt{\left( n_{+}{M}_{++}^{s} -n_{-}{M}_{--}^{s} \right) ^{2}+4\tilde{n}_{-}\tilde{n}_{+}\left( {M}_{+-}^{s} \right)
^{2}}, \\
\lambda _{4}^{\left( s\right) } &=&\frac{1}{2}\left( n_{+}{M}_{++}^{s} 
+n_{-}{M}_{--}^{s} \right)-\frac{1}{2}\sqrt{\left( n_{+}{M}_{++}^{s} -n_{-}{M}_{--}^{s} \right) ^{2}+4\tilde{n}_{-}\tilde{n}_{+}\left( {M}_{+-}^{s} \right)
^{2}}.
\end{eqnarray*}
that we shall use bellow. We have omitted the explicit dependence on $T_{m}$. 
The reduced normalized state for the intrinsic parity is 
\begin{eqnarray}
\hat{\rho}_{1}^{\left( s\right) } &=&\mathrm{Tr}_{2} \hat{\rho}
_{12}^{\left( s\right) }  \notag \\
&=&{M}_{++}^{s} \left\vert +\right\rangle \left\langle
+\right\vert +{M}_{--}^{s} \left\vert -\right\rangle
\left\langle -\right\vert +{M}_{+-}^{s} \cos \left( 2\chi
\right) \left( \left\vert +\right\rangle \left\langle -\right\vert
+\left\vert -\right\rangle \left\langle +\right\vert \right)  \label{red1} 
\end{eqnarray}
where the nondiagonal term stands for the transition probabilities $\left( \left\vert
+\right\rangle \rightleftarrows \left\vert -\right\rangle \right)$, $\left\vert
\left\langle \pm \right\vert \hat{\rho}_{1}^{\left( s\right)}\left\vert \mp
\right\rangle \right\vert ^{2}=\left( {M}_{+-}^{s}\left( T_{m}\right)
\right) ^{2}\cos ^{2}\left( 2\chi \right)$,  
so the strength of a transition depends on the mixing angle $\chi$. As the Fermi gas cools, 
$\lim_{T_m \longrightarrow 0}{M}_{+-}^{s}\left( T_{m}\right) \longrightarrow 0\,$
, independently of the value of $s$ and $\chi $, so the state decoheres
at a lower rate than it takes for the system to reduce the negative 
($ M_{--}^{1} (T_{m})$) or positive parity ($ M_{++}^{0} (T_{m})$) contribution to the mixture. 
For $ T_{m}=10^{-2}$ we find ${M}_{--}^{1} \left( T_{m}\right) /
M_{+-}^{1}\left( T_{m}\right) \approx 10^{-2}$. 
The eigenvalues of intrinsic parity state (\ref{red1}) are 
\begin{equation*}
\lambda _{\pm }^{\left( s\right) }=\frac{1}{2}\pm \frac{1}{2}\sqrt{\left( {M}
_{++}^{s}\left( T_{m}\right) -{M}_{--}^{s}\left( T_{m}\right) \right)
^{2}+\left( 2{M}_{+-}^{s}\left( T_{m}\right) \cos \left( 2\chi \right)
\right) ^{2}}.
\end{equation*}

By its turn, the normalized density operator for the helicity is diagonal 
\begin{equation*}
\hat{\rho}_{2}^{\left( s\right) }=\mathrm{Tr}_{1}\hat{\rho}_{12}^{\left(
s\right) }=H_{++}^{s}\left\vert \uparrow \right\rangle \left\langle
\uparrow \right\vert +H_{--}^{s}\left\vert \downarrow \right\rangle
\left\langle \downarrow \right\vert
\end{equation*}
and the coefficients are 
\begin{equation}
H_{++}^{s}=\frac{1}{2}+\frac{\pi }{8}\sin \left( 2\chi \right) \cos {(\mu )}
\left( {M}_{++}^{s}\left( T_{m}\right) -{M}_{--}^{s}\left( T_{m}\right)
\right) ,
\end{equation}
and
\begin{equation}
H_{--}^{s}=\frac{1}{2}-\frac{\pi }{8}\sin \left( 2\chi \right) \cos {(\mu
)}\left( {M}_{++}^{s}\left( T_{m}\right) -{M}_{--}^{s}\left( T_{m}\right)
\right) .
\end{equation}

For $\chi =\frac{n\pi }{2}$ or $\mu =\left( m+\frac{1}{2}\right) \pi $, $
n,m=0,1,2,...$, $H_{++}^{s}=$ $H_{--}^{s}=1/2$, therefore, there is no natural 
preference for any helicity direction, neither a dependence on the temperature. 
For $\chi =\frac{\pi }{4}$ or $\mu = 0$, and $s=1$, the probabilities are
unbalanced and show a dependence on the temperature, $H_{++}^{1}>H_{--}^{1}$,
since at low temperatures $\left\vert {M}_{++}^{1}\left( T_{m}\right) -{M}
_{--}^{1}\left( T_{m}\right) \right\vert \approx 1$, see Table II. In the
present epoch the helicity does not show any directional preference for the
fermions, they are found in positive and negative helicity equally likely, 
so ${\mu =\pi /2}$ is the most plausible choice for any value for the mixture 
angle $\chi $ and temperature $T_{m}$. 

\begin{figure}[tbp]
{\normalsize \includegraphics{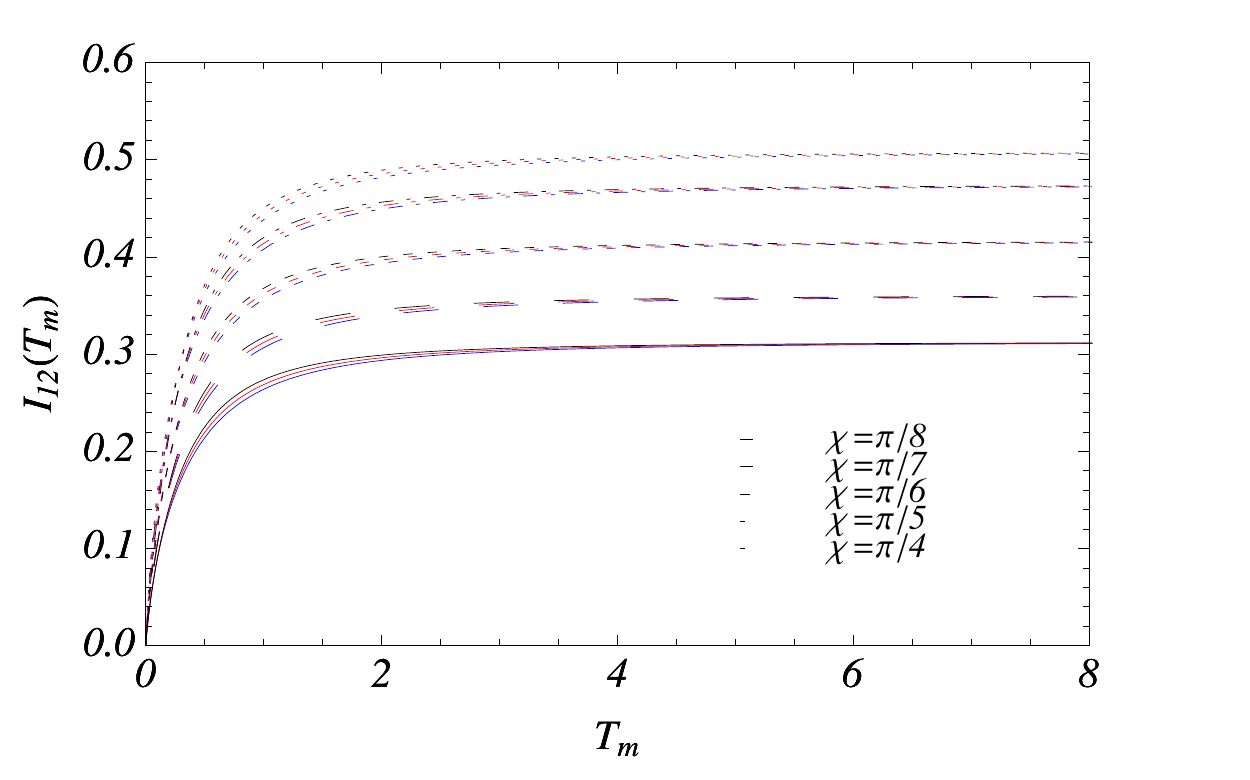} }
\caption{Mutual information $I_{12}(T_{m})$ as a function of the temperature 
parameter $T_{m}$. The plots are for several choices of the mixing angle 
$\chi $. Different lines represent different phases: $\mu =0$ (black), 
$\pi /4$(red), and $\pi /2$ (blue). For each $\chi$ and for different $\mu$, 
the curves show very tiny mutual deviations from each other. 
$I_{12}(T_{m})$ decreases to zero at the non-relativistic 
($T_{m}\rightarrow 0$) limit and increases to its maximal value at the 
ultra-relativistic ($T_{m}\gg 1$) regime.}
\label{Fig02}
\end{figure}

The von-Neumann entropy of a density operator $\hat{\rho}$ is
defined as $\mathcal{H}(\hat{\rho})=-\sum_{j}{k_{j}\,\ln {(k_{j})}}$, where $
k_{j}$ are the eigenvalues, thus we calculate the mutual information between
intrinsic parity and helicity as 
\begin{equation}
I_{12}(T_{m})=\mathcal{H}(\hat{\rho}_{1}^{\left( s\right) })+\mathcal{H}( 
\hat{\rho}_{2}^{\left( s\right) })-\mathcal{H}(\hat{\rho}_{12}^{\left(s
\right) }),
\end{equation}
for several values of the mixture angles $\chi $ and the phase $\mu $,
as depicted in Fig. \ref{Fig02}. 

The variation of $\mu $ implies into some very tiny mutual
deviations from each other at the transition regime (from UR to NR), i.e.,
$\mu $ has not any relevant qualitative effect onto the mutual information between parity 
and helicity, thus reinforcing our previous hint for the choice $\mu = 0$. 
As expected, the mutual information $I_{12}(T_{m})$ is insignificant at low temperature, $
T_{m}\rightarrow 0$, while it is maximal for $T_{m}\gg 1$. 

\section{Summary and conclusions}

It is common knowledge \cite{kolb90} that at the very early universe 
($T \geq 10^{12} \ K$) the photons had enough energy to become
electron-positron pairs, so electrons and positrons existed in thermal
equilibrium with the radiation. At about 400 000 years after the Big Bang
there is change, radiation is free to pass through the universe as its expansion
changes it from opaque to transparent. As the universe expanded it cooled, and when
the temperature reduced to $\approx 10^{9} \ K$ photons had not enough energy to
create $e^{-}-e^{+}$ pairs, so electrons and positrons were no longer in
thermal equilibrium but radiation acquired a thermalized blackbody distribution. 
A fundamental question is: why matter (positive
intrinsic parity) eventually dominated over anti-matter (negative intrinsic
parity), which, presumably, were initially in equal footing? Our calculations 
and results cannot explain the ``disappearance'' of the antifermions that existed in
the early universe, however it hints of why at the present epoch the
fermions -- the quarks and leptons that constitute matter -- have positive intrinsic 
parity (our estimate is $10^{20}$ positive for 1 negative parity fermions) and any
produced antifermion has a negative intrinsic parity, although the 
calculations show that at the early universe fermions and antifermions existed 
in a superposition of both parities entangled the helicity states. 
Otherwise, on the non-relativistic limit the mutual information is null: 
any quantum correlation between the particle/antiparticle character 
and the state spin-polarization vanishes.
It corresponds to an issue that can be reproduced, from the mathematical point of view, 
by a Foldy-Wouthuysen unitary transformation \cite{Zub80}]. 

Finally, we point out the essentiality of the present framework 
\cite{salomon, salomon2}, where it was assumed that the Dirac equation 
and the spinors describe the dynamics and the state of
a massive particle carrying two qubits of information, the helicity and the intrinsic parity.
That approach permits quantifying the quantum correlation and the entanglement between the 
particle/antiparticle degrees of freedom. Moreover, we believe that it might be relevant 
discussing the destruction of the ``mirror'' symmetry (external parity 
or \emph{left/right}-handed character) in particle decays involving electroweak 
interactions, a point that certainly deserves to be scrutinized in the 
subsequent investigations. 

\section{Acknowledgments}
AEB acknowledges financial support from CNPq (grant 300809/2013-1). SSM acknowledges 
financial support from CNPq and from INCT-IQ.


\begin{thebibliography}{99}
\bibitem{nota2} In this context, it 
is worth mentioning that the photon can also be cast in
terms of a quantum wave equation $i\hslash \partial \boldsymbol{F}/\partial
t=c\boldsymbol{\vec{s}}\cdot \vec{p}\boldsymbol{F}$, where $\boldsymbol{F}$
is a spinor having three components, and their real and imaginary part\ are
the electric and magnetic fields respectively. $\boldsymbol{\vec{s}}$
components are the spin 1 matrices and $\vec{p}$ is the linear momentum.
This approach for the photon was studied in \cite{iwo}.
\end{thebibliography}
\end{document}